\documentclass[10pt,journal,compsoc]{IEEEtran}

\IEEEoverridecommandlockouts
\usepackage{amsmath,amssymb,amsfonts}

\usepackage[table]{xcolor}
\usepackage{cleveref}
\usepackage{subfig}
\usepackage{graphicx}
\usepackage{url}

\definecolor{myGreen}{RGB}{143,188,143}
\definecolor{myYellow}{RGB}{246,239,110}
\definecolor{myRed}{RGB}{180,92,92}

\newcommand\phase[1]{\tikz[baseline=(X.base)]\node [draw=black,fill=white,thick,rectangle,inner sep=2pt, rounded corners=2pt](X){\color{black}\textbf{#1}};}

\usepackage{soul}
\def\BibTeX{{\rm B\kern-.05em{\sc i\kern-.025em b}\kern-.08em
    T\kern-.1667em\lower.7ex\hbox{E}\kern-.125emX}}
    
\usepackage{xspace}    
\usepackage{booktabs}
\newcommand{\simulink}{Simulink\textsuperscript{\tiny\textregistered}\xspace}
\newcommand{\Stateflow}{Stateflow\textsuperscript{\tiny\textregistered}\xspace}

\newcommand{\SLDV}{Simulink\textsuperscript{\tiny\textregistered} 
Design Verifier\textsuperscript{\tiny TM}\xspace}
\newcommand{\SLTest}{Simulink\textsuperscript{\tiny\textregistered}
Test\textsuperscript{\tiny TM}\xspace}
\newcommand{\Simscape}{Simscape\textsuperscript{\tiny\textregistered}\xspace}
\newcommand{\SimscapeElectrical}{Simscape\textsuperscript{\tiny\textregistered}
Electrical\textsuperscript{\tiny TM}\xspace}
\newcommand{\SimscapeDriveline}{Simscape\textsuperscript{\tiny\textregistered}
Driveline\textsuperscript{\tiny TM}\xspace}

\usepackage{algorithm}
\usepackage{algorithmic}

\usepackage{amsmath}

\usepackage{tcolorbox}
\newtcolorbox{mybox}{colback=red!5!white,colframe=red!75!black}

\newcommand{\NAME}{\textsc{HECATE}\xspace}

\newcommand{\foresee}{\textsc{ForeSee}\xspace}

\newcommand{\falsify}{\textsc{falsify}\xspace}

\newcommand{\FalCAuN}{\textsc{FalCAuN}\xspace}
\newcommand{\staliro}{\textsc{S-Taliro}\xspace}
\newcommand{\Aristeo}{\textsc{Aristeo}\xspace}

\newcommand{\Breach}{\textsc{Breach}\xspace}
\newcommand{\FalStar}{\textsc{FalStar}\xspace}

\definecolor{keywordcolor}{HTML}{1e46be}
\definecolor{stepcolor}{HTML}{724722}

\newcommand{\lit}[1]{\textbf{\texttt{\textcolor{keywordcolor}{#1}}}}

\newcommand{\testassessmenttofitnesscalculator}{\textsc{TA2FF}\xspace}
\newcommand{\simulinkvariable}[1]{\text{#1}}
\newcommand{\model}{\textsc{M}}
\newcommand{\testsequence}{\textsc{TS}\xspace}
\newcommand{\failingtestsequence}{\textsc{TC}}
\newcommand{\candidatetestsequence}{\ensuremath{\testsequence\_\textsc{c}}\xspace}

\newcommand{\testassessment}{\textsc{TA}\xspace}
\newcommand{\searchspace}{\textsc{SP}}
\newcommand{\fitnessfunction}{\textsc{FF}}
\newcommand{\fitnessvalue}{\textsc{f}}
\newcommand{\fitnesscalculator}{\textsc{FitnessFunction}\xspace}

\newcommand{\fitnessconverter}{\text{FITNESS\_CONVERTER}\xspace}
\newcommand{\aggregator}{\text{AGGREGATOR}\xspace}

\newcommand{\numsimulinkmodels}{\ensuremath{16}\xspace}

\newcommand{\numberOfTestsForEachModelBenchmark}{300\xspace}

\newcommand{\totalNumberOfTestsForEachModelBenchmark}{6000\xspace}

\newcommand{\numModelsDifferentVerdicts}{two\xspace}

\newcommand{\numModelsDifferentVerdictsTranslation}{one\xspace}
\newcommand{\numModelsDifferentVerdictsRequirement}{one\xspace}

\newcommand{\nruns}{\ensuremath{50}\xspace}
\newcommand{\RQonenumberofDays}{15\xspace}

\newcommand{\RQonehecatevsstaliro}{26.1\%\xspace}

\newcommand{\RQonehecatevsstalirobetterpercentage}{81\xspace}
\newcommand{\RQonehecatevsstalirobettermodels}{13\xspace}

\newcommand{\RQonehecatevsstaliroequalpercentage}{13\xspace}
\newcommand{\RQonehecatevsstaliroequalmodels}{2\xspace}

\newcommand{\RQonehecatevsstalirolowerpercentage}{6\xspace}
\newcommand{\RQonehecatevsstalirolowermodels}{1\xspace}

\newcommand{\RQonehecatevsstaliropreferredpercentage}{$\approx$94\%\xspace}
\newcommand{\RQonehecatevsstaliropreferredmodels}{15\xspace}

\newcommand{\RQonehecateavg}{98.1\%\xspace}
\newcommand{\RQonehecatemin}{70.0\%\xspace}
\newcommand{\RQonehecatemax}{100.0\%\xspace}
\newcommand{\RQonehecatestd}{7.5\%\xspace}

\newcommand{\RQonestaliroavg}{72.0\%\xspace}
\newcommand{\RQonestaliromin}{4.0\%\xspace}
\newcommand{\RQonestaliromax}{100.0\%\xspace}
\newcommand{\RQonestalirostd}{30.5\%\xspace}

\newcommand{\RQtwohecateavg}{28.4\xspace}
\newcommand{\RQtwohecatemin}{4.2\xspace}
\newcommand{\RQtwohecatemax}{134.9\xspace}
\newcommand{\RQtwohecatestd}{31.2\xspace}

\newcommand{\RQtwostaliroavg}{99.9\xspace}
\newcommand{\RQtwostaliromin}{3.4\xspace}
\newcommand{\RQtwostaliromax}{276.0\xspace}
\newcommand{\RQtwostalirostd}{72.8\xspace}

\newcommand{\RQtwohecateavgtime}{148.8s\xspace}
\newcommand{\RQtwohecatemintime}{3.7s\xspace}
\newcommand{\RQtwohecatemaxtime}{1503.6s\xspace}
\newcommand{\RQtwohecatestdtime}{369.2s\xspace}

\newcommand{\RQtwostaliroavgtime}{816.6s\xspace}
\newcommand{\RQtwostaliromintime}{2.1s\xspace}
\newcommand{\RQtwostaliromaxtime}{5255.6s\xspace}
\newcommand{\RQtwostalirostdtime}{1571.4s\xspace}

\newcommand{\RQtwohecatemoreefficient}{$\approx$81\%\xspace}
\newcommand{\RQtwohecatevsstaliro}{71.5\xspace}
\newcommand{\RQtwohecatevsstalirotime}{667.8s\xspace}

\newcommand{\TrueValue}{\textsc{true}\xspace}
\newcommand{\FalseValue}{\textsc{false}\xspace}
\newcommand{\nff}{\textsc{NFF}\xspace}

\usepackage[framemethod=TikZ]{mdframed}
\newenvironment{Answer}[1][]{
  \ifstrempty{#1}
  {\mdfsetup{
    frametitle={
      \tikz[baseline=(current bounding box.east),outer sep=0pt]
      \node[line width=0pt,anchor=east,rectangle,draw=white,fill=white]
    ;}}
  }
  {\mdfsetup{
    frametitle={
      \tikz[baseline=(current bounding box.east),outer sep=0pt]
      \node[anchor=east,rectangle,draw=white,fill=white]
    {\strut #1};}}
  }
  \mdfsetup{innertopmargin=-5pt,linecolor=black,
            linewidth=0.5pt,topline=true,
            frametitleaboveskip=\dimexpr-\ht\strutbox\relax,}
  \begin{mdframed}[]\relax
  }{\end{mdframed}}

\newcommand{\stepAAIModethree}{\textbf{\textbf{\color{stepcolor}AAI\_Mode\_3}}\xspace}
\newcommand{\stepAAIModethreeoff}{\textbf{\textbf{\color{stepcolor}AAI\_Mode\_3\_OFF}}\xspace}
\newcommand{\stepAAIModethreeon}{\textbf{\color{stepcolor}AAI\_Mode\_3\_ON}\xspace}
\newcommand{\stepEnd}{\textbf{\color{stepcolor}END}\xspace}

\newcommand{\stepselection}{\textbf{\color{stepcolor}Mode\_selection}\xspace}
\newcommand{\stepmodethree}{\textbf{\color{stepcolor}Mode\_3}\xspace}
\newcommand{\stepsensing}{\textbf{\color{stepcolor}Sensing}\xspace}
\newcommand{\stepheartbeat}{\textbf{\color{stepcolor}Heartbeat}\xspace}
\newcommand{\stepnoheartbeat}{\textbf{\color{stepcolor}No\_heartbeat}\xspace}

\newcommand{\stepelse}{\textbf{\color{stepcolor}Else}\xspace}

\usepackage[switch]{lineno}

\newcommand{\linebreakand}{
  \end{@IEEEauthorhalign}
  \hfill\mbox{}\par
  \mbox{}\hfill\begin{@IEEEauthorhalign}
}

\begin{document}

\title{Simulation-based 
Testing of Simulink Models with Test Sequence and Test Assessment Blocks
}

\author{Federico~Formica,
Tony~Fan,
Akshay~Rajhans,
Vera~Pantelic,
Mark~Lawford, 
Claudio~Menghi
\thanks{F. Formica, T.~Fan, V.~Pantelic, M.~Lawford and C.~Menghi are 
with the McMaster University, Hamilton, Canada -
 e-mail: \{formicaf,fant6,pantelv,lawford, menghic\}@mcmaster.ca\newline
A. Rajhans is with MathWorks, MA, USA - email:  arajhans@mathworks.com
}
}

\maketitle

\begin{abstract}
Simulation-based software testing supports engineers in finding faults in  \simulink models.
It typically relies on search algorithms that iteratively generate test inputs used to exercise models in simulation to detect design errors. 
While simulation-based software testing techniques are effective in many practical scenarios, they are typically not fully integrated within the Simulink environment and require additional manual effort.
Many techniques require engineers to specify requirements using logical languages that are neither intuitive nor fully supported by Simulink, thereby limiting their adoption in industry.

This work presents \NAME, a testing approach for Simulink models using Test Sequence and Test Assessment blocks from \SLTest.
Unlike existing testing techniques,  \NAME uses information from Simulink  models to guide the search-based exploration. 
Specifically, \NAME relies on information provided by the Test Sequence and Test Assessment blocks to guide the search procedure.
Across a benchmark of $\numsimulinkmodels$ Simulink models from different domains and industries, our comparison of \NAME with the state-of-the-art testing tool \staliro indicates that \NAME is both more effective (more failure-revealing test cases) and efficient (less iterations and computational time) than \staliro for \RQonehecatevsstaliropreferredpercentage and \RQtwohecatemoreefficient of benchmark models respectively. 
Furthermore, \NAME successfully generated a failure-revealing test case for a representative case study from the automotive domain demonstrating its practical usefulness. \end{abstract}


\IEEEpeerreviewmaketitle

\section{Introduction}
\label{sec:intro}
Engineers often use \simulink~\cite{Simulink} to design and demonstrate software behavior~\cite{elberzhager2013analysis,DBLP:journals/sosym/LiebelMTLH18}.
It offers a graphical programming environment to simulate and generate code from  graphical models.
Simulink is used across industries including  medical, avionics, and automotive~\cite{dajsuren2013simulink,boll2021characteristics}.
Techniques that support \emph{verification of Simulink models}, a significant and widely recognized software engineering problem~\cite{Aristeo,menghi2019generating}, are of interest to academia and industry alike.
There are many techniques to verify Simulink models including
formal methods and simulation-based software testing~\cite{kapinski2016simulation}.

\emph{Formal methods} tools such as \SLDV~\cite{SimulinkDesignVerifier}, FRET~\cite{giannakopoulou2020generation}, and QVtrace~\cite{8569614} support verification of Simulink models.
For example,  Simulink Design Verifier 
uses 
formal verification for 
detection of design errors 
(e.g., division-by-zero), identification of dead logic, 
generation of test cases to ensure some amount of test coverage, 
and exhaustive verification of properties capturing system requirements. 
Formal methods tools typically have constraints on the kinds of Simulink models that can be analyzed~\cite{10.1145/3338906.3340444}. 
For example, Simulink Design Verifier runs a compatibility check 
before its execution~\cite{ProveProperties} and cannot perform verification on incompatible models. 
In contrast, simulation-based software testing, the technique considered by this work, does not impose restrictions on the kinds of Simulink models that can be analyzed.

\emph{Simulation-Based Software Testing}, a widely used technique to test Simulink models~\cite{kapinski2016simulation}, relies on simulations to detect 
software failures.
Simulation test cases are either manually specified by the user  (e.g.,~\cite{arrieta2020seeding,schmidt2016model,rajhans2021specification}) or are automatically generated 
(e.g.,~\cite{Aristeo,fainekos2019robustness,9155827,10.1007/978-3-030-81685-8_29,10.1007/978-3-030-76384-8_24,matinnejad2018test,schmidt2014efficient,Reactis}).

\emph{Manual test case specification} requires engineers to manually specify test inputs and assessment procedures to verify whether requirements are satisfied or violated. This approach is largely used in the industrial domain~\cite{juhnke2021challenges} and is mandated by safety standards~\cite{ISO26262}. 
Manual test case specification is supported in \SLTest~\cite{SimulinkTest} via the \emph{Test Sequence} block that lets users construct complex test inputs~\cite{TestSequence}, and the \emph{Test Assessment} block that lets users specify the procedure to check the requirements of interest~\cite{TestAssessment}. 
Manual test case specification builds on human capabilities (e.g., domain knowledge) to define test cases.
However, the manual definition of test cases is a laborious task especially for large-scale industrial projects~\cite{schmidt2014efficient}. 
Therefore, there is a need for automated support to alleviate some of the manual effort.

\emph{Automated test case generation} entails automatic 
generation of test cases~\cite{kapinski2016simulation}. 
For example, falsification-based testing tools for Simulink models 
iteratively explore the input space to detect failure-revealing test cases (e.g.~\cite{Aristeo,fainekos2019robustness,9155827,10.1007/978-3-030-81685-8_29,10.1007/978-3-030-76384-8_24}).
Falsification-based testing  tools for Simulink models include 
\Aristeo~\cite{Aristeo},
\Breach~\cite{Breach}, 
\FalCAuN~\cite{Waga20},
\falsify~\cite{yamagata2020falsification}, 
\FalStar~\cite{ErnstSZH2018-FalStar},
\foresee~\cite{falsQBRobCAV2021}, and
\staliro~\cite{S-Taliro}.
However, these tools are not based on native Simulink Test blocks that are often used by industry practitioners. 
They also require engineers to specify requirements and input profiles using formal languages (e.g., Signal Temporal Logic~\cite{maler2004monitoring})---a time-consuming and error-prone task, as widely recognized by the software engineering community~\cite{czepa2019modeling,czepa2018understandability}. This limitation makes the application of falsification-based techniques laborious and 
hinders their use in industrial applications.
For this reason, this paper proposes an automated test case generation technique that supports Test Sequence and Test Assessment blocks (together referred to as \emph{Test Blocks} for brevity in the rest of the paper).

To summarize, the problem addressed by this work is to \emph{support automation of test case generation from manual test case specifications done using Simulink Test Sequence and Test Assessment blocks}.

To address this problem, we present \NAME, a Test Sequence and Test Assessment driven Approach for Simulink models.
\NAME bridges the gap between manual test case specification and automated test case generation.
Specifically, \NAME is the first simulation-based software testing  approach that supports Test Blocks. 
On the one hand, unlike existing automated test case generation techniques, \NAME leverages manual test case specification artifacts (Test Blocks) to guide the search-based exploration. On the other hand, unlike existing manual test case generation techniques, \NAME also systematically automates the generation of test cases. 

\NAME requires engineers to extend their test sequences into Parameterized Test Sequences: test sequences augmented with a set of parameters and their domains. 
These Parameterized Test Sequences define the universe of test sequences to be considered during the search.
\NAME automatically searches for parameters values that lead to test sequences that show a violation of the conditions specified by the Test Assessment block.
To enable the search, \NAME provides a procedure that automatically converts Test Assessment blocks into fitness functions.
\NAME is implemented as a plugin for \staliro~\cite{S-Taliro}, a widely known testing tool for Simulink models.

To evaluate effectiveness and efficiency of \NAME, we identify (a)~a baseline tool, and (b)~a benchmark for our comparison.
For the baseline tool, 
we use \staliro.
Unlike \NAME, \staliro requires Signal Temporal Logic (STL)~\cite{maler2004monitoring} specifications and input profiles to describe the requirement to be considered and the search domain (together referred to STL artifacts in the rest of the paper).
As part of the contribution of this work, we propose a new benchmark of  \numsimulinkmodels Simulink models that includes both Test Blocks and STL artifacts.
The benchmark includes models from an international competition on falsification-based testing tools~\cite{DBLP:conf/arch/ErnstABCDFFG0KM21}, from the web (\cite{simulinkpacemaker,TestSequence, benchmarkHPS,benchmarkST,benchmarkFS,benchmarkTL}), and includes models developed by Toyota~\cite{jin2014powertrain} and Lockheed Martin~\cite{mavridou2020ten,benchmarkLM}.
Our results show that for our benchmark models, \NAME is more effective than \staliro in finding failure-revealing test cases for \RQonehecatevsstaliropreferredpercentage of the models.
\NAME was also more efficient than \staliro for \RQtwohecatemoreefficient of the models. 
To evaluate the applicability of \NAME,
we selected a large and representative automotive case study developed by MathWorks 
in the context of the EcoCAR Mobility Challenge~\cite{EcoCAR}, a competition they jointly organize with the U.S.\ Department of Energy
and General Motors.
We check whether \NAME could generate a failure-revealing test case.
\NAME successfully returned a failure-revealing test case for our case study.

To summarize, our contributions are as follows:
\begin{itemize}
    \item \NAME, a testing approach for Simulink models with Test Blocks (\Cref{sec:contribution});
    \item the use of Parameterized Test Sequences (\Cref{sec:tstoIp});
    \item compilation of Test Assessment blocks into fitness functions (\Cref{sec:tatoFit});
    \item a benchmark of \numsimulinkmodels Simulink models that include both Test Blocks and STL artifacts (\Cref{sec:benchmark});
    \item effectiveness and efficiency assessment of \NAME by comparing it with \staliro (\Cref{sec:rq1} and \Cref{sec:rq2});
    \item applicability assessment of \NAME on a representative case study from the automotive industry (\Cref{sec:rq3});
    \item and a complete replication package.
\end{itemize}

The structure of this paper is as follows:
Section~\ref{sec:pacemaker} introduces Test Sequence and Test Assessment blocks.
Section~\ref{sec:contribution} presents \NAME.
Section~\ref{sec:SimulinkDriver} and~\ref{sec:searchEngine} describe the two main phases of \NAME: the driver and the search phases. 
Section~\ref{sec:evaluation} evaluates our contribution.
Section~\ref{sec:discussion} discusses our results and threats to validity.
Section~\ref{sec:related} presents related work.
Section~\ref{sec:conclusion} concludes summarizing our results. 

 \newcommand{\TU}{\simulinkvariable{TU}}

\begin{figure}[t]
    \includegraphics[width=\columnwidth]{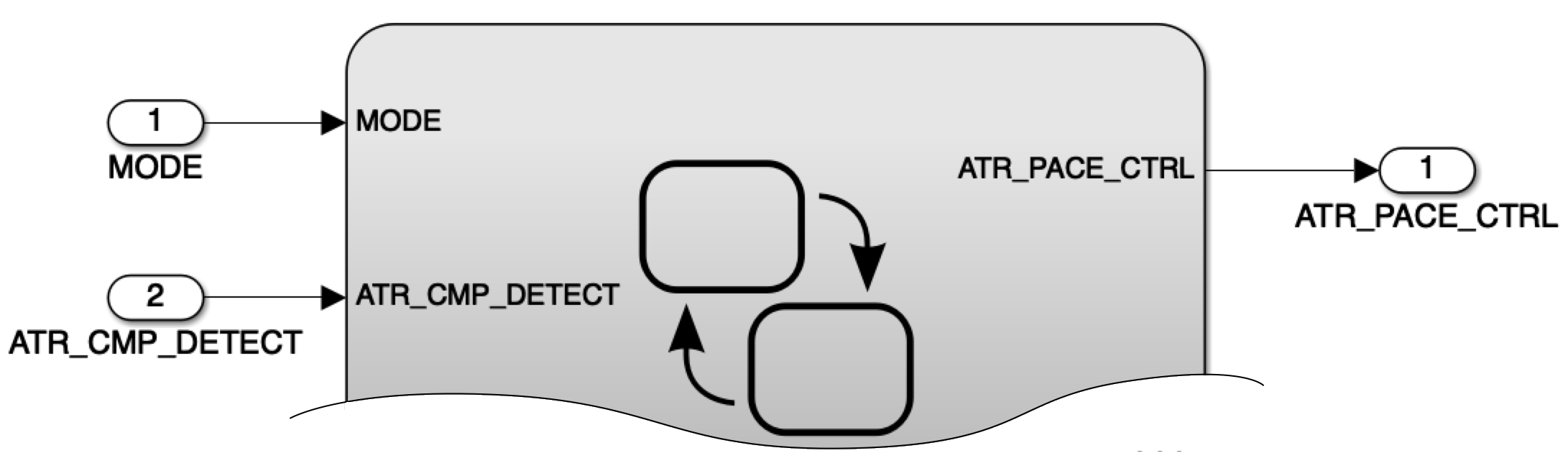}
    \caption{A portion of the \simulink model of a pacemaker controller.
}
    \label{fig:pacemakerSimulink}
\end{figure}

\section{Background and Running Example}
\label{sec:pacemaker}
This section describes an illustrative example of a pacemaker model~\cite{simulinkpacemaker} to informally introduce the syntax and semantics of Test Blocks.

A pacemaker is a medical device that regulates its user's heart rate. Its control software can be designed using Simulink. 
\Cref{fig:pacemakerSimulink} depicts a portion of the controller model.
The pacemaker is connected to the heart atrium and ventricle using sensors to check heart activity and actuators to deliver electrical signals for muscle activation.
The portion of the Simulink subsystem (gray box) shown in \Cref{fig:pacemakerSimulink} includes input ports \simulinkvariable{MODE} and \simulinkvariable{ATR\_CMP\_DETECT} and an output port \simulinkvariable{ATR\_PACE\_CTRL}. Input signals from these input ports are  the desired operational mode for the pacemaker and 
a signal that specifies how the atrial pulse from the heart is detected, respectively.
The output signal to the output port indicates whether the atrium should be paced.

\begin{figure}[t]
 \subfloat[Test Sequence\label{fig:testSequence}]{
    \centering
    \includegraphics[width=\columnwidth]{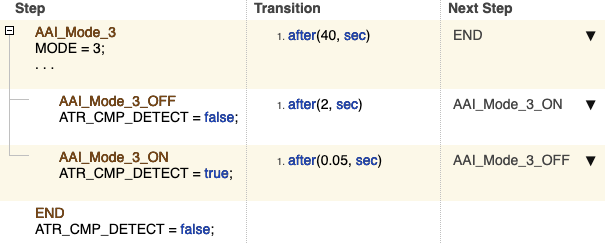}}\newline
 \subfloat[Test Assessment\label{fig:testAssessment}]{
    \centering
    \includegraphics[width=\columnwidth]{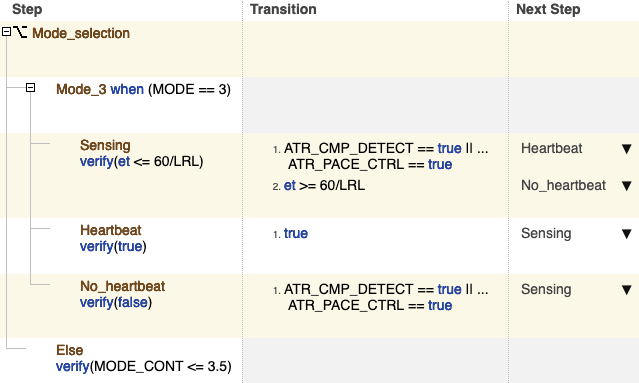}}
    \caption{\simulink Test Sequence and Test Assessment blocks for the pacemaker example. 
    }
    \label{fig:pacemaker}
\end{figure}

\Cref{fig:pacemaker} presents portions of Test Sequence and Test Assessment blocks for the pacemaker example that introduce the syntax and semantics of these two block types~\cite{TSSyntax}.
Test Blocks are comprised of \emph{test steps} connected by \emph{transitions}.
The Test Sequence block in \Cref{fig:testSequence} and the Test Assessment block in \Cref{fig:testAssessment} contains four (\stepAAIModethree, \stepAAIModethreeoff, \stepAAIModethreeon, \stepEnd) and six (\stepselection, \stepmodethree, \stepsensing, \stepheartbeat, \stepnoheartbeat, \stepelse) test steps respectively.
Test steps are hierarchically organized. 
For example, the test step
\stepsensing of \Cref{fig:testAssessment} is nested within the test step
\stepmodethree. 
At each time instant, the Test Blocks are in exactly one `leaf' test step and its `ancestors' in the hierarchy.
For example, if the Test Assessment in \Cref{fig:testAssessment} is in the leaf test step \stepsensing, it is also in all of its ancestors, namely \stepmodethree and
\stepselection. Whenever a Test Block enters a `non-leaf' test step, it also enters its first child test step. 
For example, if the
\stepmodethree test step of \Cref{fig:testAssessment} is entered, the test step
\stepsensing is also entered.

Transitions define how a Test Block switches between test steps. 
For example, transitions of the Test Assessment block from \Cref{fig:testAssessment} specify how the Test Assessment switches between the test steps
\stepselection,
\stepmodethree,
\stepsensing,
\stepheartbeat,
\stepnoheartbeat,
and 
\stepelse. 
Whenever a non-leaf test step is exited, so are all of its `descendants'. 
For example, if the Test Assessment of \Cref{fig:testAssessment} is in the test step \stepsensing and the value of the \simulinkvariable{MODE} changes to the value~$2$,
the \stepmodethree test step is left and the test step \stepsensing is also left.
Two precedence rules regulate the firing of transitions:
First, if two transitions associated with test steps at the different levels of the hierarchy are active at the same time instant, the transition at the highest level of the hierarchy is fired.
Second, if two transitions associated with the same test step become active, the first among the defined transitions is fired.
Transitions are classified as \emph{standard transitions} and \emph{when decompositions}.

\emph{Standard transitions} connect a source and a destination test step.
They are labelled with a Boolean formula representing a condition 
for the transition to be fired.
The formula is defined using transition operators, temporal operators, and relational operators summarized next (see~\cite{TSSyntax} for
additional information).

\begin{enumerate}
    \item \emph{Transition operators} 
    return Boolean values and evaluate the occurrence of signal events.
    For example, the transition operator \lit{hasChanged}(\simulinkvariable{u}) detects if signal \simulinkvariable{u} changed its value compared to the previous timestep.
    
    \item \emph{Temporal operators} 
 return  
    values providing timing information related to a Test Block. 
    For example, the temporal operators \lit{after}(\simulinkvariable{n}, \TU) and \lit{et}(\TU) respectively evaluates if \simulinkvariable{n} units of time have elapsed since the beginning of the current test step, and retrieves the time elapsed from since a test step is entered.
    
    \item \emph{Relational operators} ($<,\leq,>,\geq,=$) compare the values of arithmetic expressions defined on the values assumed by signal inputs, parameters, or constants.
    Arithmetic expressions can be defined by using   arithmetic and temporal operators.
    
\end{enumerate}
As an example,  the standard transition from the test step \stepAAIModethree to the test step
\stepEnd in \Cref{fig:testSequence} is associated with the condition \lit{after}(40, \textbf{\color{keywordcolor}sec}) which becomes true after the system has been in the test step \stepAAIModethree for $40$s.

\emph{When decomposition} corresponds to switch statements from programming languages. 
For example, the when decomposition of the Test Assessment block in \Cref{fig:testAssessment} specifies that the Test Block is in the test step
\stepmodethree
 depending on whether the value assumed by the variable 
 \simulinkvariable{MODE} is $3$.

\begin{figure}[t]
 \subfloat[Test Input\label{fig:testinput}]{
    \centering
    \hspace{0.6cm}
    \includegraphics[width=0.74\columnwidth]{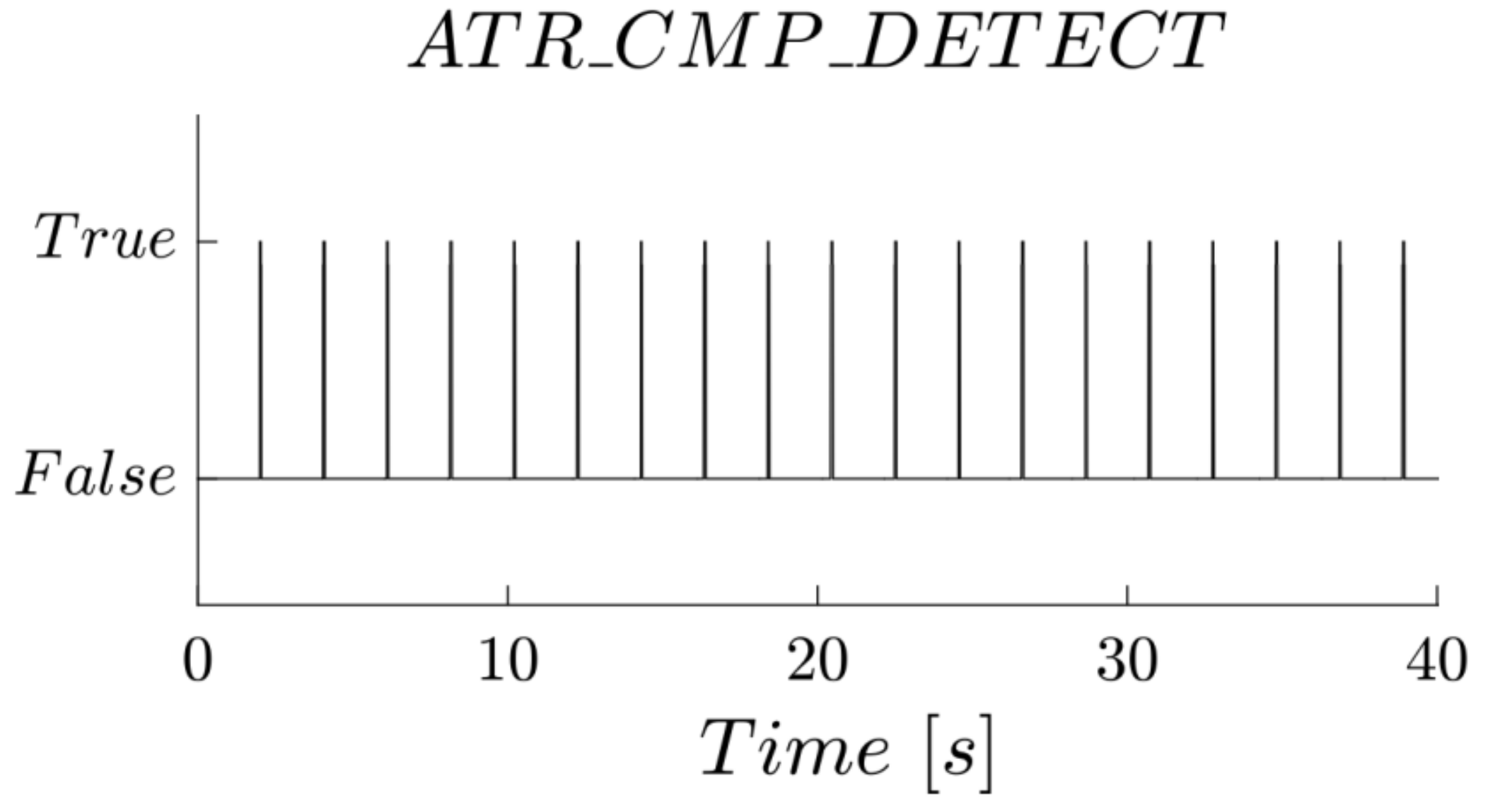}    \hspace{0.9cm}
      }\newline
 \subfloat[Test Output\label{fig:testresult}]{
    \centering
        \hspace{0.1cm}
    \includegraphics[width=0.8\columnwidth]{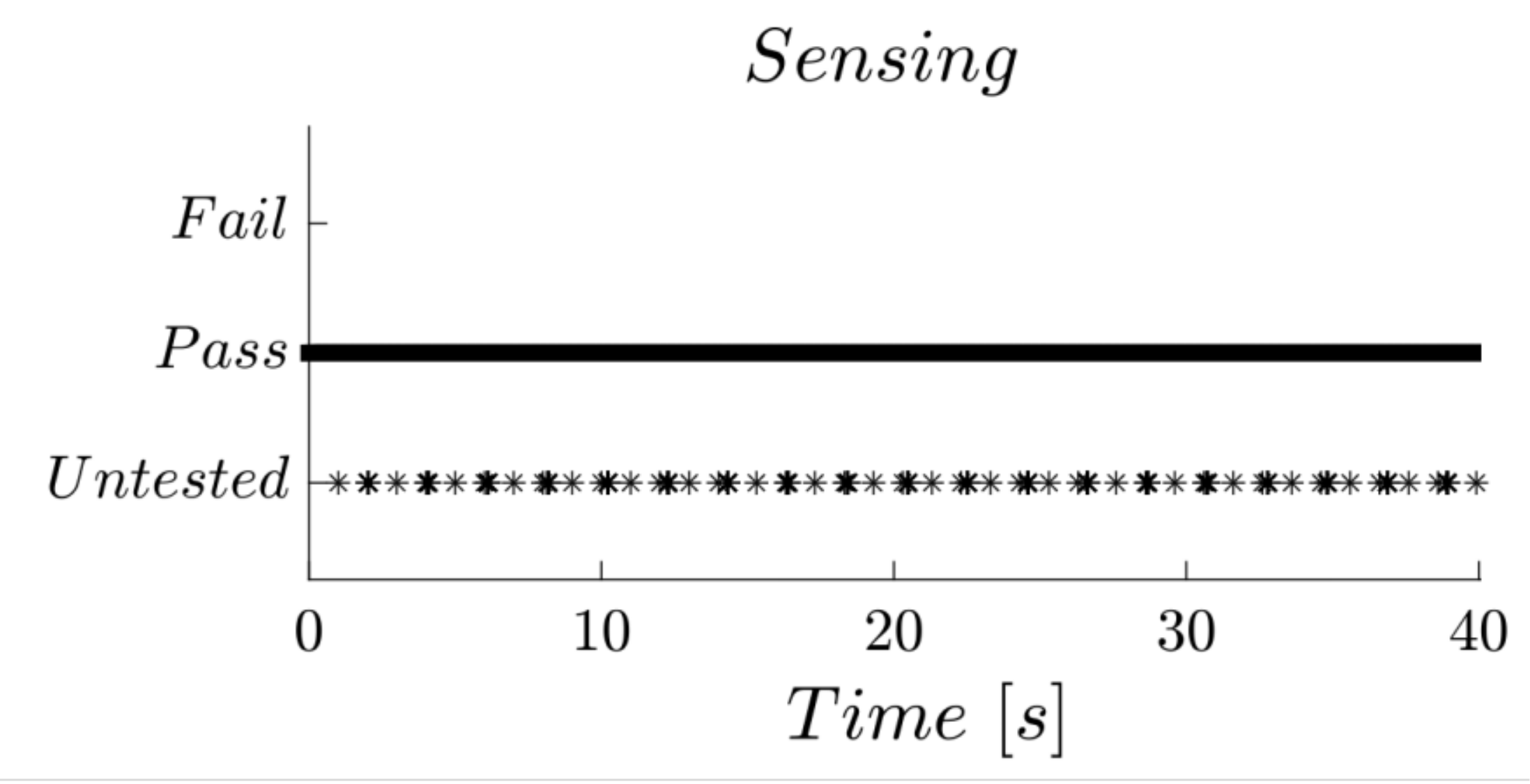}    \hspace{0.9cm}
    }
    \caption{Input signal generated by the Test Sequence of \Cref{fig:testSequence} and output signal of the Test Assessment of \Cref{fig:testAssessment}. 
}
    \label{fig:inputandresult}
\end{figure}

Test steps from Test Sequence and Test Assessment blocks differ in their purpose in the following manner.
In Test Sequences, they define the values assumed by the input signals. For example, the test steps \stepAAIModethreeoff
 and \stepAAIModethreeon
 of the Test Sequence of \Cref{fig:testSequence} assign the values \lit{false} and \lit{true} to the variable \simulinkvariable{ATR\_CMP\_DETECT}, respectively. 
In contrast, in Test Assessment, they contain verification statements, i.e., \lit{verify} and \lit{assert}, which check whether a logical expression evaluates to \TrueValue  or \FalseValue.
For example, the statement \lit{verify}(\lit{et}$<$60/\simulinkvariable{LRL}) 
of the test step \stepsensing 
in \Cref{fig:testAssessment} checks whether or not the pacemaker spends less than 60/\simulinkvariable{LRL} seconds sensing for a heartbeat. 
The lower rate limit \simulinkvariable{LRL}  changes depending on the user activity, e.g., sleeping, running. 
Differently from the \lit{verify} statement, 
the \lit{assert} statement  also stops the simulation if the logical expression  evaluates to \FalseValue. 

A \emph{test case} is made of a Test Sequence block and a Test Assessment block. 
For example, the Test Sequence and Test Assessment blocks in \Cref{fig:pacemaker} form a test case for the Simulink model of \Cref{fig:pacemakerSimulink}.

The Test Sequence block generates a test input containing one input signal for each input port. 
For example, \Cref{fig:testinput} presents an input signal for the input port \simulinkvariable{ATR\_CMP\_DETECT} from the Test Sequence block in \Cref{fig:testSequence}.
The input signal \simulinkvariable{ATR\_CMP\_DETECT} of the test input reported in \Cref{fig:testinput} repeatedly changes its value between \TrueValue and \FalseValue following the specification of the Test Sequence block from \Cref{fig:testSequence}.
More specifically, the Test Sequence block switches from
\stepAAIModethreeoff
 to 
\stepAAIModethreeon
 every $2$s, and vice versa every $0.05$s. 
Therefore, the value of the signal  \simulinkvariable{ATR\_CMP\_DETECT} (\Cref{fig:testinput}) switches between \FalseValue to \TrueValue every $2$s, and vice versa every $0.05$s.
The test input is then executed by simulating the model for the input signals associated with the Test Sequence. 

Test Assessment block evaluates if the output signals of the model lead to a violation of the expressions of its \lit{verify} and \lit{assert} statements.
It produces an output signal for each \lit{verify} and \lit{assert} statement.
For example, \Cref{fig:testresult} presents the output signal produced by the \lit{verify} statement of the
\stepsensing  test step from the Test Assessment block in \Cref{fig:testAssessment}.
For each time instant, the value of the output signal is computed as follows.
If the Test Assessment is not in the \stepsensing test step, the output signal is marked \simulinkvariable{Untested}. Otherwise it is assigned a value of \simulinkvariable{Pass} or \simulinkvariable{Fail}, depending on whether the logical expression contained within the  \lit{verify} statement is \TrueValue or \FalseValue.
In our example, the value for this statement is always \simulinkvariable{Pass} since the Test Assessment either moves to the test step \stepheartbeat or 
\stepnoheartbeat before the condition \lit{et}$<$60/\simulinkvariable{LRL} evaluates to \FalseValue. 
A verification statement (\lit{verify} or \lit{assert}) is violated if its logical expression evaluates to \FalseValue at least once, i.e., the output signal is \FalseValue for at least one simulation step.
A test case (Test Sequence plus Test Assessment) is considered a \emph{failure-revealing test case} for the model if it violates a verification statement.

Despite the large support provided by existing tools for testing Simulink models (see \Cref{sec:intro}), none of them can automatically generate Simulink test cases from manual test case specifications defined using Test Sequence and Test Assessment blocks.
This lack of support is a significant limitation for practical applications.
In the next section, we present \NAME, a systematic framework that overcomes this limitation by using manual test case specification to automatically generate new failure-revealing test cases.

\tikzstyle{output} = [coordinate]
\begin{figure}[t]
\centering
\begin{tikzpicture}[auto,
 block/.style ={rectangle, draw=black, thick, fill=white!20, text width=5em,align=center, rounded corners},
 block1/.style ={rectangle, draw=blue, thick, fill=blue!20, text width=5em,align=center, rounded corners, minimum height=2em},
 line/.style ={draw, thick, -latex',shorten >=2pt},
 cloud/.style ={draw=red, thick, ellipse,fill=red!20,
 minimum height=1em}]

\draw (0,0) node[block] (Input) {\phase{3} \footnotesize Input \\ Generation};
\node [output, right of=Input,node distance=1.7cm] (InputMiddle) {};
 \node [block, below of=InputMiddle,node distance=0.3cm,text width=6.5cm,minimum height=2.1cm,dashed,draw opacity=1] (Search) {};
 
 \draw (0,0) node[block] (InputRef) {\phase{3} \footnotesize Input \\ Generation};
 
  \node [block, above of=InputMiddle,node distance=2.4cm,text width=6.5cm,minimum height=2.05cm,dashed] (Driver) {};

\node[block, right of=Input,node distance=3.5cm] (Fitness){\phase{4} \footnotesize Fitness\\ 
Assessment};

\node [output, right of=Fitness,node distance=2.5cm] (uoutput) {};
 \node [block, above of=Fitness,node distance=2.2cm,text width=2.2cm] (TestAssessment) {\phase{2} \footnotesize Test Assessment to
 Fitness Function};
 \node [block, above of=Input,node distance=2.2cm,text width=2.2cm] (TestToInput) {\phase{1}\\ 
 \footnotesize Definition of the
 Search Space
};
\node [output, above of=TestToInput,node distance=1.7cm] (testsequence) {};
\node [output, above of=TestAssessment,node distance=1.7cm] (testassessment) {};
\node [output, below of=Fitness,node distance=1cm] (loopa) {};
\node [output, below of=Input,node distance=1cm] (loopb) {};
\draw (0,0) node [above of=Driver, node distance=0.8cm] (DriverText) {\footnotesize \textbf{Driver Phase}};
\draw (0,0) node [below of=Search, node distance=0.9cm] (SearchText) {\footnotesize \textbf{Search Phase}};
\node [output, right of=Fitness,node distance=0.5cm] (modela) {};

\node [output, below of=modela,node distance=1.8cm] (model) {};
\node [output, below of=modela,node distance=0.5cm] (modelb) {};

\draw[-stealth] (model.east) -- (modelb.south)    node[
pos=0.2,
right]{\model};
\draw[-stealth] (Input.east) -- (Fitness.west)    node[midway,above]{\candidatetestsequence};
\draw[-stealth] (testsequence.south) -- (TestToInput.north)
    node[near start,right]{\testsequence};
\draw[-stealth] (TestToInput.south) -- (Input.north)
    node[midway,right]{\searchspace};
\draw[-stealth] (testassessment.south) -- (TestAssessment.north)
    node[near start,right]{\testassessment};
\draw[-stealth] (TestAssessment.south) -- (Fitness.north)
    node[midway,right]{\fitnessfunction};    
\draw[-] (Fitness.south) -- (loopa.north)
    node[midway,right]{};    
\draw[-] (loopb.south) -- (loopa.north)
    node[midway,above]{$\langle \testsequence_1, \fitnessvalue_1 \rangle$,$\langle \testsequence_2, \fitnessvalue_2 \rangle$, \ldots};
\draw[-stealth] (loopb.north) -- (Input.south)
    node[midway,right]{};    
 \draw[-stealth] (Fitness.east) -- (uoutput.west) node[midway,above]{\failingtestsequence/\nff};
 \end{tikzpicture}
\caption{Overview of \NAME.}
\label{fig:contribution}
\end{figure}

\section{Test Sequence and Assessment Driven Testing}
\label{sec:contribution}

\newcommand{\codespace}{\hspace{0.5cm}}

\newcommand{\inputgeneration}{\texttt{SearchDomainDefinition}}

Figure~\ref{fig:contribution} presents an overview of \NAME.
The inputs are the Test Sequence (\testsequence) and Test Assessment (\testassessment) blocks and the Simulink model under test (\model). 
The output is either a failure-revealing test case (\failingtestsequence) or the value \nff (No Fault Found) indicating that \NAME could not detect a failure-revealing test case within the allotted time budget.
\NAME is composed of a \emph{driver} phase and a \emph{search} phase.

The \emph{driver} phase compiles the 
Test Blocks into artifacts that drive the simulation-based search in two steps.
Step \phase{1} defines the search space (\searchspace) associated with the Test Sequence block.
Step \phase{2} translates the Test Assessment block into a fitness function (\fitnessfunction) that guides the search. 

The \emph{search} phase implements the iterative testing procedure of the simulation-based software testing framework in two steps.
Step~\phase{3} iteratively generates new candidate Test Sequences (\candidatetestsequence) using the search space definition (\searchspace).
Step~\phase{4} executes the model (\model) for the test inputs specified by the candidate Test Sequence. 
It checks using the fitness function (\fitnessfunction) whether the Test Assessment is satisfied or violated.

The tool stops either by returning the failure-revealing test case (\failingtestsequence) if a violation is detected, or if the computation exceeds its allotted time budget (\nff).  
Otherwise, it proceeds with a new iteration and step~\phase{3} uses the previously generated test sequences and their fitness values (i.e., $\langle \testsequence_1, \fitnessvalue_1 \rangle$,$\langle \testsequence_2, \fitnessvalue_2 \rangle$) to  generate a new candidate Test Sequence (\candidatetestsequence).

Driver and Search Phases are detailed in Sections~\ref{sec:SimulinkDriver} and~\ref{sec:searchEngine}.

\section{Driver Phase}
\label{sec:SimulinkDriver}
This section describes the two steps of the driver phase:
Definition of the Search Space (\Cref{sec:tatoFit}) and Test Assessment to Fitness Function Conversion (\Cref{sec:tstoIp}).

\subsection{Definition of the Search Space}
\label{sec:tstoIp}
In step \phase{1} users amend the Test Sequence to define the search space, i.e., the universe of test sequences to be considered by  \NAME for test case generation. 
Specifically, \NAME requires engineers to (a)~create a Parameterized Test Sequence,
and (b)~define domains for the parameters.

\begin{figure}
    \subfloat[Parameterized Test Sequence \label{fig:TestSequenceParam}]{
    \centering
    \includegraphics[width=\linewidth]{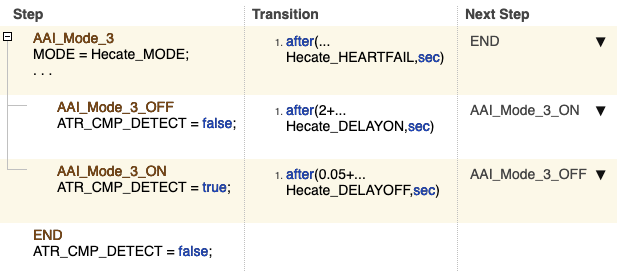}
    }
    \vspace{0.5cm}\newline
    \subfloat[Parameter Domain\label{fig:InputParam}]{
    \centering
    \includegraphics[width=0.8\linewidth]{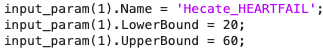}
    }
        \vspace{0.2cm}\newline
    \caption{Parameterized Test Sequence  (\Cref{fig:TestSequenceParam}) and parameter domain (\Cref{fig:InputParam}) for the Test Sequence from \Cref{fig:testSequence}.}
    \label{fig:Para,eterizedTestSequence}
\end{figure}

\emph{Creating a Parameterized Test Sequence}. 
To create a Parameterized Test Sequence engineers extend a Test Sequence by manually introducing search parameters. 
Search parameters are  variables named with identifiers starting with the string ``\simulinkvariable{Hecate}'' that can be used as Simulink variables.
For example, \Cref{fig:TestSequenceParam} reports a Parameterized Test Sequence  with search parameters
 \simulinkvariable{Hecate\_HEARTFAIL}, \simulinkvariable{Hecate\_DELAYON} and \simulinkvariable{Hecate\_DELAYOFF} for the Test Sequence from \Cref{fig:testSequence}. 

\emph{Definition of parameter domains}.
Software engineers define parameter domains by providing lower and upper bound values.
For example, \Cref{fig:InputParam} sets the values $20$ and $60$ as lower and upper bounds for the parameter \simulinkvariable{Hecate\_HEARTFAIL}.

Parameterized Test Sequences and parameter domains define the search space (\searchspace) of the testing algorithm and are provided as input to the search phase.

\subsection{Test Assessment to Fitness Function Conversion}
\label{sec:tatoFit}
Step~\phase{2} converts the Test Assessment into a fitness function (\fitnessfunction)  that guides the search procedure.
The fitness function needs to ensure the following: 
(i)~a negative fitness value indicates that the property is violated;
(ii)~a positive fitness value indicates that the property is satisfied; 
(iii)~the higher the positive fitness value is, the farther the system is from  violating the property; and 
(iv)~the lower the negative value, the farther the system is from satisfying the property.

\begin{figure}
\subfloat[Model of the \text{FITNESS\_FUNCTION} subsystem~(\fitnessfunction).
    \label{fig:FitCalculator}]{
    \centering
    \includegraphics[width=1\linewidth]{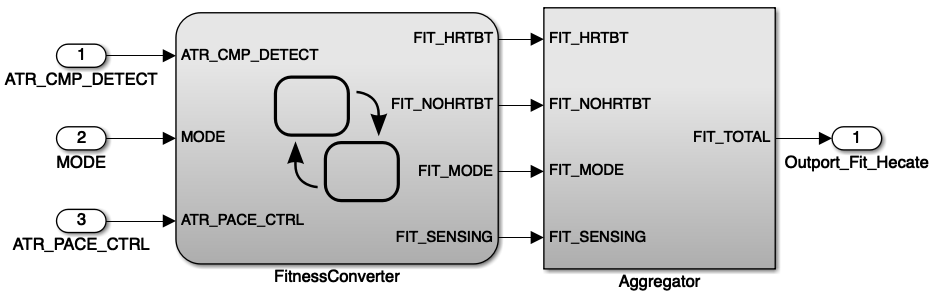}}\vspace{0.5cm}\newline
    \subfloat[\fitnessconverter subsystem (\fitnessfunction).
    \label{fig:SFChart}]{
    \centering
\includegraphics[width=\linewidth]{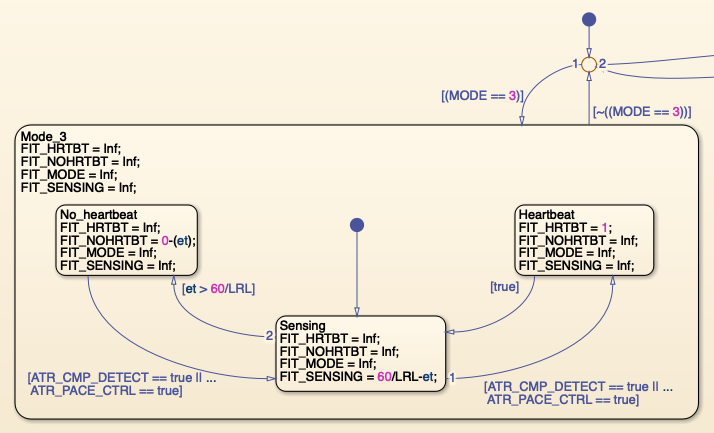} 
    } \vspace{0.5cm}\newline
    \subfloat[\aggregator subsystem (\textsc{AG}).
    \label{fig:FitConverter}]
    {
    \vspace{0.2cm}\centering
    \includegraphics[width=\linewidth]{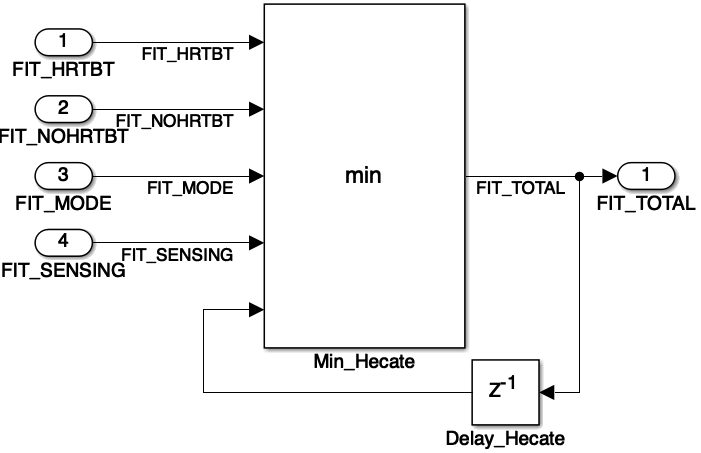}}
        \vspace{0.2cm}\newline\caption{Fitness Function generated for the Test Assessment block from \Cref{fig:testAssessment}.
        }
    \label{fig:TestSequenceUpdate}
\end{figure}

The function \testassessmenttofitnesscalculator~(Test Assessment to fitness function) from \Cref{alg:hecate} implements Step~\phase{2}.
It takes a Test Assessment block (\testassessment) as input and returns a Simulink subsystem (\fitnessfunction) representing the fitness function.
For example, \Cref{fig:TestSequenceUpdate} presents the fitness function generated for the Test Assessment block from \Cref{fig:testAssessment}.
The \testassessmenttofitnesscalculator algorithm proceeds as follows.
Line~\ref{alg:createsFC} creates the \fitnesscalculator (\fitnessfunction) subsystem (\Cref{fig:FitCalculator}). 
Line~\ref{alg:createsSF} creates its two subsystems: the \fitnessconverter (\textsc{FC})  and the \aggregator (\textsc{AG}).
The \fitnessconverter  has the same input ports as the Test Assessment block, and one output port for each verification statement (\lit{verify} and \lit{assert}). 
Each output port returns the fitness value associated to the corresponding verification statement.
For example, the subsystem \fitnessconverter in \Cref{fig:FitCalculator} has four output ports, one for each of the verification statements from \Cref{fig:testAssessment}.
The output port \simulinkvariable{FIT\_SENSING} is associated to the statement
\lit{verify}(\lit{et}$<$60/\simulinkvariable{LRL}).
The \aggregator subsystem computes a single fitness value from the fitness values of the output ports of the \fitnessconverter subsystem.

\begin{figure}[t]
\begin{algorithmic}[1]
\STATE \textbf{function} \testassessmenttofitnesscalculator(\testassessment)
\STATE \label{alg:createsFC} \codespace \fitnessfunction=\texttt{createFitnessFunction}();\\
\STATE \label{alg:createsFCV} \codespace [\textsc{FC},\textsc{AG}]=\texttt{initialize}(\fitnessfunction);\\
\STATE \label{alg:createsSF} \codespace \textsc{FC}.\texttt{createStateflow}();\\
\STATE \label{alg:statesCreation} \codespace \textsc{FC}.\texttt{createStates}(\testassessment);\\
\STATE \label{alg:AddTransitions} \codespace \textsc{FC}.\texttt{createTransitions}(\testassessment);\\
\STATE \label{alg:FitnessConversion} \codespace \textsc{FC}.\texttt{addFitnessComputation}(\textsc{TA});\\
\STATE \label{alg:min} \codespace  \textsc{AG}.\texttt{addMinCalculation}(\textsc{SS});
\STATE \label{alg:addFeedback} \codespace  \textsc{AG}.\texttt{addFeedback}(\textsc{SS});
\STATE \codespace \textbf{return}  \fitnessfunction;
\STATE \textbf{end}
\end{algorithmic}
\caption{The \testassessmenttofitnesscalculator algorithm.}
\label{alg:hecate}
\end{figure}

To generate the \fitnessconverter (\Cref{fig:SFChart}) Algorithm~\ref{alg:hecate} 
creates a \Stateflow chart~\cite{Stateflow} that mimics the behaviors of the test steps.
Line~\ref{alg:createsSF} creates an empty Stateflow chart. 
Line~\ref{alg:statesCreation} creates the states of the Stateflow chart: one for each test step of the \testassessment nested according to the test steps as necessary. 
For example, the state \simulinkvariable{No\_heartbeat} of the Stateflow chart in \Cref{fig:SFChart} corresponds to the test step
\stepnoheartbeat from \Cref{fig:testAssessment} and is nested within the state \simulinkvariable{Mode\_3} as the test step  of the Test Assessment block is nested within the test step
\stepmodethree. 
Line~\ref{alg:AddTransitions} adds the transitions to the \fitnessconverter chart by analyzing each transition of the Test Assessment block.
If the transition is a standard transition, the algorithm adds a transition to the Stateflow chart of the  \fitnessconverter with the same condition that connects the Stateflow states associated with the source and destination test steps of the transition of the Test Assessment.
For example, the transition of the Test Assessment in \Cref{fig:testAssessment} connecting the test step
\stepnoheartbeat  to \stepsensing translates into the transition of the Stateflow chart from \Cref{fig:SFChart} connecting the state \simulinkvariable{No\_heartbeat} to \simulinkvariable{Sensing}.
If a transition is a when decomposition, 
the algorithm adds to the Stateflow chart a junction node 
and two transitions for each  \lit{when} statement: one transition connecting the junction node to the Stateflow state corresponding to the test step of the Test Sequence that is the target of the \lit{when} statement, and another transition with the opposite source and destination states.
The first transition is guarded with the condition of the \lit{when} statement. 
The second is associated with the negation of that condition or the satisfaction of the \lit{when} statement of any of the previous states.
Lastly, 
Line~\ref{alg:FitnessConversion} translates the arithmetic expression within each  verification statement into a fitness metric. 
We use the robustness~\cite{fainekos2006robustness}  fitness metric since it is widely used for search-based testing (e.g.,~\cite{fainekos2019robustness,10.1007/978-3-030-81685-8_29,Aristeo}).

To generate the \aggregator subsystem (\Cref{fig:FitConverter}) Algorithm~\ref{alg:hecate} proceeds as follows.
Line~\ref{alg:min} adds a \texttt{min} Simulink block to the \aggregator subsystem that computes the minimum value assumed by the signals of its input ports.
For example, 
the  \texttt{Min} block of
the \aggregator subsystem (\Cref{fig:FitConverter}) generated for the Test Assessment from \Cref{fig:testAssessment} is connected to the input ports \simulinkvariable{FIT\_HRTBT}, 
\simulinkvariable{FIT\_NOHRTBT},
\simulinkvariable{FIT\_MODE}, and 
\simulinkvariable{FIT\_SENSING}.
Finally, Line~\ref{alg:addFeedback} creates a feedback loop with a delay block (to capture the minimum value at the previous time) to compute the minimum fitness value over the entire simulation.

\section{Search Phase}
\label{sec:searchEngine}
This section describes the two steps of the search phase:
Input Generation  (Section~\ref{sec:inputGeneration}) and Fitness Assessment (Section~\ref{sec:fitnessAssessment}).

\subsection{Input Generation}
\label{sec:inputGeneration}
Input Generation (step \phase{3}) takes as inputs the search space definition (\searchspace) as well as the Test Sequences previously generated by \NAME ($\testsequence_1$, $\testsequence_2$, \ldots) and their fitness values (\fitnessvalue$_1$, \fitnessvalue$_2$, \ldots). 
It generates a new candidate Test Sequence (\candidatetestsequence) by assigning values to the parameters of the Parameterized Test Sequence within their domains.
For example, the Test Sequence from \Cref{fig:testSequence} is obtained by assigning the value~0 to all its parameters from the Parameterized Test Sequence from \Cref{fig:TestSequenceParam}.

\NAME reuses  \staliro~\cite{S-Taliro} to generate parameters values.
We selected  \staliro since it offers a set of search algorithms, such as Simulated Annealing~\cite{abbas2014robustness}, Monte Carlo~\cite{nghiem2010monte}, and gradient descent methods~\cite{abbas2014functional}.
\NAME assigns parameters values computed by \staliro to the Parameterized Test Sequence.
For example, the Test Sequence presented in \Cref{fig:testSequence} is a possible Test Sequence returned by \NAME.

\subsection{Fitness Assessment}
\label{sec:fitnessAssessment}
The Fitness Assessment (step \phase{4}) computes the fitness value for a Test Sequence.
This step executes the Simulink model augmented with the 
\text{FITNESS\_FUNCTION} subsystem by invoking the \texttt{sim} command~\cite{sim}. 
It then extracts the fitness value for the Test Sequence as the last value assumed by the output signal \simulinkvariable{FIT\_TOTAL}.

 \section{Evaluation}
\label{sec:evaluation}

This section empirically evaluates \NAME by answering  the following research questions:
\begin{itemize}
    \item[RQ1] How \emph{effective} is \NAME in generating failure-revealing test cases? (Section~\ref{sec:rq1})
    \item[RQ2] How \emph{efficient} is \NAME in generating failure-revealing test cases? (Section~\ref{sec:rq2})
    \item[RQ3]  How \emph{useful} is \NAME in generating failure-revealing test cases for a large automotive model? (Section~\ref{sec:rq3})
\end{itemize}

To answer RQ1 and RQ2, we compared \NAME with a  baseline framework from the literature.
Since we are not aware of any other testing techniques that consider Test Blocks, we had to select a testing framework that does not support them.
We considered a framework that supports STL artifacts, i.e., input profiles and STL specifications (see~\cite{Aristeo} for a detailed description), since they are widely used by existing testing frameworks and considered by international tool competitions~\cite{DBLP:conf/arch/ErnstABCDFFG0KM21,ARCHWEBSITE}.
Among the available tools,
our baseline is \staliro since it was classified as ready for industrial development~\cite{kapinski2016simulation}, and already used in several domains (e.g.,~\cite{tuncali2018experience}). 
To enable our comparison,
we had to define a new benchmark 
since we are not aware of any existing benchmark containing Simulink models with both Test Blocks and STL artifacts. Our benchmark (see \Cref{sec:benchmark})  is one of the contributions of this work.
We compared the effectiveness and efficiency of \NAME and \staliro in finding failure-revealing test cases for our benchmark models. 
Notice that a  failure-revealing test case for \NAME consists of a Test Sequence and a Test Assessment block,
whereas for \staliro it consists of an input profile and an STL specification. 

To answer RQ3, we assessed whether \NAME could find failure-revealing test cases for a large representative automotive case study.

\emph{Implementation}.
\NAME is a plugin for \staliro. 
A complete replication package containing our models, data, and the \NAME tool is publicly available~\cite{HECATETool}.
The EKF model is protected by an non-disclosure agreement and can not be shared publicly.

\begin{table*}[t]
    \centering
    \caption{Identifier (MID), description, number of blocks (\#Blk), input ports (\#In), output ports (\#Out),
    simulation time in seconds (Ts),  percentage of failure-revealing runs  for \NAME and \staliro  (RQ1), and time and number of iterations --- within parentheses ---  for \NAME and \staliro  (RQ2)
   for our benchmark models.
    }
    \label{tab:models}
    \footnotesize
    \begin{tabular}{l l r r r r  |r r| r r}
        \toprule
        & & & & & &\multicolumn{2}{c|}{\textbf{RQ1}} &\multicolumn{2}{c}{\textbf{RQ2}} \\
        \textbf{MID} &\textbf{Description} &\textbf{\#Blk}   &\textbf{\#In}   &\textbf{\#Out}  
        & \textbf{Ts}
        & \textbf{\NAME} & \textbf{\staliro} & \textbf{\NAME} & \textbf{\staliro} \\
        \midrule
        AFC     & Controller for the air-to-fuel ration in an engine.  &309    &2  &1  
        &$50$   & \cellcolor{myGreen!50}100 \% & \cellcolor{myGreen!50}60 \%    & \cellcolor{myGreen!50}1503.6 (52.8)  & \cellcolor{myGreen!50}5255.6 (134.8)\\
        AT      & Car automatic transmission with gears.   &79   &2   &3 
        &$20$   & \cellcolor{myGreen!50}100 \% & \cellcolor{myGreen!50}4 \% & \cellcolor{myGreen!50}29.9 (33.4)  & \cellcolor{myGreen!50}588.1 (276.0)\\
        CC      & Controller for a system formed by five cars.   &26 &2  &5  
        &$100$  & \cellcolor{myRed!50}70 \% & \cellcolor{myRed!50}74 \%  & \cellcolor{myRed!50}265.9 (134.9)  & \cellcolor{myRed!50}317.5 (130.4)\\
        NN      & Neural Network controller for a magnet.   &139    &1 &2 
        &$5$    & \cellcolor{myGreen!50}100 \% & \cellcolor{myGreen!50}56 \%  & \cellcolor{myGreen!50}13.3 (27.4)  & \cellcolor{myGreen!50}160.0 (154.5)\\
        SC      &Dynamic model of steam condenser.   &184    &1 &4  
        &$35$   & \cellcolor{myGreen!50} 100 \% & \cellcolor{myGreen!50}18 \% & \cellcolor{myGreen!50}29.1 (34.3)  & \cellcolor{myGreen!50} 355.8 (219.4)\\
        WT      &  Wind turbine considering wind speed.   &177    &1  &6  
        &$630$  &\cellcolor{myGreen!50}100 \% &\cellcolor{myGreen!50}90 \%    & \cellcolor{myGreen!50}127.1 (30.7)  & \cellcolor{myGreen!50}269.8 (51.0)\\
        AT2     &Different version of the AT model. &80 &2  &2 
        &$45$   &\cellcolor{myGreen!50}100 \%    &\cellcolor{myGreen!50}74 \% & \cellcolor{myGreen!50}4.1 (7.6)   & \cellcolor{myGreen!50}74.4 (122.5)\\
        EKF     & Car battery with an Extended Kalman Filter.  &263  &1  &1  
        &$9000$  &\cellcolor{myGreen!50}100 \%    &\cellcolor{myGreen!50}90 \% & \cellcolor{myGreen!50}9.1 (6.6)   & \cellcolor{myGreen!50}151.2 (46.2)\\
        FS      &  Damping system for wing oscillations.   &440   &2   &2  
        &$15.01$ &\cellcolor{myGreen!50}100 \% &\cellcolor{myGreen!50}98 \%    & \cellcolor{myGreen!50}66.2 (19.4)  & \cellcolor{myGreen!50}222.3 (68.5)\\
        HPS     & Heat pump controlling the room temperature.  &54 &2  &5  
        &$300$   & \cellcolor{myGreen!50}100 \% & \cellcolor{myGreen!50}30 \%  & \cellcolor{myRed!50}3.7 (6.9)   & \cellcolor{myRed!50}2.1 (3.4)\\
        PM      & Controller for a Pacemaker. &62 &2  &4  
        &$60$   &\cellcolor{myGreen!50} 100 \% & \cellcolor{myGreen!50}88 \%  & \cellcolor{myGreen!50} 27.2 (25.9)  & \cellcolor{myGreen!50}4306.1 (69.0)\\
        ST      & Signal tracker with 3 available tracking modes.  &30 &2  &3  
        &$30$   &\cellcolor{myYellow!50}100 \%    &\cellcolor{myYellow!50}100 \%  & \cellcolor{myRed!50}34.7 (25.9)  & \cellcolor{myRed!50}26.3 (24.6)\\
        TL      & Traffic model of a crossroad.    &84 &2  &2  
        &$1000$ & \cellcolor{myGreen!50}100 \% & \cellcolor{myGreen!50}92 \%  & \cellcolor{myGreen!50}203.3 (15.6)  & \cellcolor{myGreen!50}840.9 (67.2)\\
        TUI     & Tustin Integrator for flight control.    &74  &2  &1  
        &$20$   &\cellcolor{myYellow!50}100 \%    &\cellcolor{myYellow!50}100 \%  & \cellcolor{myGreen!50}17.9 (4.2)   & \cellcolor{myGreen!50}121.5 (31.3)\\
        NNP   & Neural Network Predictor with  hidden layers.     &714    &2  &1  
        &$100$  &\cellcolor{myGreen!50}100 \%    &\cellcolor{myGreen!50}96 \% & \cellcolor{myGreen!50}30.0 (19.7)    & \cellcolor{myGreen!50}141.9 (83.7)\\
        EU      & Computes the Rotation matrix  of Euler angles.  &174    &6  &3  
        &$10$   & \cellcolor{myGreen!50} 100 \% & \cellcolor{myGreen!50} 82 \%    & \cellcolor{myGreen!50}15.4 (9.8)   & \cellcolor{myGreen!50}232.2 (115.2)\\
    \bottomrule
    \end{tabular}
\end{table*}

\subsection{Definition of our Benchmark}
\label{sec:benchmark}
Our benchmark contains \numsimulinkmodels Simulink models:
six models are from the ARCH competition~\cite{DBLP:conf/arch/ErnstABCDFFG0KM21} (AFC, AT, CC, NN, SC, WT),
seven are from the web (AT2~\cite{TestSequence}, EKF~\cite{9813961}, FS~\cite{benchmarkFS}, HPS~\cite{benchmarkHPS}, PM~\cite{simulinkpacemaker}, ST~\cite{benchmarkST},  TL~\cite{benchmarkTL}),
and three are from a benchmark developed by Lockheed Martin (EU, NNP, TUI)~\cite{mavridou2020ten,benchmarkLM}.
These models either come only with STL artifacts (AFC, AT, CC, NN, SC, WT), or only Test Blocks (AT2, FS, HPS, PM, ST, TL), or neither (EKF, EU, NNP, TUI).
For each model, we considered one testing scenario (Test Blocks and STL artifact).
For example, for each model of the ARCH competition we considered one STL artifact, i.e., a combination made by an STL specification and an input profile (AT-1 for AT, AFC29 for AFC, NNX for NN, WT2 for WT, CC1 for CC, SC for SC --- see~\cite{DBLP:conf/arch/ErnstABCDFFG0KM21} for further details).
To ensure that each model comes with both Test Blocks and STL artifact, we manually designed Test Blocks for the STL artifact and vice versa. For models in which neither the STL artifact nor the Test Blocks were available, we defined both the Test Blocks and the STL artifact by consulting their online documentation.
We introduced a fault in the models that did not violate their requirements.
Models, Test Blocks, and STL artifacts are included in our replication package.

We tested whether the Test Assessments and oracles generated from the STL specifications returned the same verdict for the same test inputs.
Specifically, for each model, we generated test inputs by reusing the procedure offered by \staliro and using a uniform distribution to generate the values of the control points inside their ranges, and pchip or piecewise-constant as the interpolation function depending on the model (see~\cite{S-Taliro} for further details on input generation).
We considered \numberOfTestsForEachModelBenchmark  test inputs (as mandated by the ARCH competition), simulated the model, and compared the verdicts from the Test Assessment and the test oracle generated from the STL specification. 
If they returned the same verdict, we considered the next model. 
Otherwise, we identified and fixed the problem, and repeated the testing process by considering the \numberOfTestsForEachModelBenchmark  test inputs previously generated together and an additional set of \numberOfTestsForEachModelBenchmark test inputs.
In total, we ran \totalNumberOfTestsForEachModelBenchmark test inputs that enabled us to detect \numModelsDifferentVerdicts problems: \numModelsDifferentVerdictsTranslation caused by bugs in our translation (\Cref{sec:tatoFit}), and \numModelsDifferentVerdictsRequirement caused by mismatch between the STL specification and  the Test Assessment block.
After fixing these problems, the verdicts were consistent for all the test inputs.

We did not compare Parameterized Test Sequences and input profiles since their goal is different. 
A Parameterized Test Sequence contains parameters representing uncertainties on the values assumed by some variables of the Test Sequence. 
Therefore, a Parameterized Test Sequence does not represent the entire domain of the input signals, but rather captures ``small'' perturbations of a Test Sequence.
Unlike Parameterized Test Sequences, input profiles represent the entire domain of the input signals.
RQ1 and RQ2 empirically assess the impact of the usage of Parameterized Test Sequences and input profiles on the effectiveness and efficiency of the testing framework.

\subsection{Effectiveness (RQ1)}
\label{sec:rq1}
To answer RQ1, we compared \NAME and \staliro  by considering simulated annealing as the search algorithm for both since it is the default algorithm for \staliro. 
We ran \NAME and \staliro for each benchmark model by setting the maximum number of search iterations to \numberOfTestsForEachModelBenchmark. 
Every run was repeated \nruns times to account for the stochastic nature of the algorithm, as done in similar works (e.g.,~\cite{Aristeo}) and mandated by the ARCH competition.
We executed experiments on a large computing platform.
For each tool, we recorded which of the \nruns runs returned a failure-revealing test case.

\textbf{Results}. 
We ran all the experiments for approximately \RQonenumberofDays days worth of total computational time, which completed in two days thanks to the parallelization facilities of our computing platform.

\Cref{tab:models} (column RQ1) reports the percentage of failure-revealing  runs  (over \nruns runs) for each benchmark model for \NAME and \staliro.
For example, for the model AT, 
\NAME and \staliro respectively returned a failure-revealing test case for $100\%$ and $4\%$ of the runs.
Cells with green, yellow, and red background in Table~\ref{tab:models} respectively refer to  cases in which the percentage of failure-revealing runs of \NAME is higher than, equal to, and lower than that of \staliro.
The percentage of failure-revealing runs of \NAME compared to that of \staliro is respectively higher for $\RQonehecatevsstalirobetterpercentage \%$ of the models ($\RQonehecatevsstalirobettermodels$ out of \numsimulinkmodels),  
equal 
for $\approx \RQonehecatevsstaliroequalpercentage \%$ of the models (\RQonehecatevsstaliroequalmodels out of \numsimulinkmodels), and  
lower 
for $\approx \RQonehecatevsstalirolowerpercentage \%$ of the models (\RQonehecatevsstalirolowermodels out of \numsimulinkmodels).
Based on these results, \NAME would be preferred over \staliro for  $\RQonehecatevsstaliropreferredmodels$ ($\RQonehecatevsstalirobettermodels+\RQonehecatevsstaliroequalmodels$) or \RQonehecatevsstaliropreferredpercentage ($\RQonehecatevsstalirobetterpercentage\%+\RQonehecatevsstaliroequalpercentage\%$) of the benchmark models where the percentage of failure-revealing runs of \NAME is higher than or equal to that of \staliro.

On average  (across the different models) \NAME and \staliro  returned a failure-revealing test case for  
\RQonehecateavg  (\textit{min}=\RQonehecatemin,
\textit{max}=\RQonehecatemax,
\textit{StdDev}=\RQonehecatestd) and
\RQonestaliroavg  (\textit{min}=\RQonestaliromin,
\textit{max}=\RQonestaliromax,
\textit{StdDev}=\RQonestalirostd) of the runs.
Therefore, on average \NAME  generated  \RQonehecatevsstaliro ($\RQonehecateavg-\RQonestaliroavg$) more  failure-revealing test cases than \staliro.
Only for CC \staliro returned  failure revealing test cases for more runs than \NAME. 
Wilcoxon rank sum test confirms that our results are statistically significant ($\alpha$=$0.05$, p=$1.67\cdot 10^{-5}$).

Therefore, \NAME effectively guided search-based exploration.
Notice that, we designed the Test Sequences to target critical areas of the input domain as commonly done in software testing and our parameterization focuses the search around these areas. 
This is a threat to the validity of our results (see \Cref{sec:discussion}).
However, the effectiveness of \NAME is likely to increase if the model developers are knowledgeable about the problem domains and thus able to better engineer Parameterized Test Sequences for their models.

\newpage
\begin{Answer}[RQ1 - Effectiveness]
\NAME is more effective than \staliro in finding failure-revealing test cases for \RQonehecatevsstaliropreferredpercentage of our benchmark models. It  generated on
average \RQonehecatevsstaliro 
more  failure-revealing test cases than \staliro.
\end{Answer}

\subsection{Efficiency (RQ2)}
\label{sec:rq2}
To answer RQ2, we compared the efficiency of \NAME and \staliro in generating failure-revealing test cases.
For each tool, we considered the runs from RQ1 that returned a failure-revealing test case and analyzed the computational time and number of iterations required to detect the failure-revealing test case.

\begin{figure}[t]
\centering
 \subfloat[Time (s)\label{fig:rq2resultstime}]{
    \centering
    \includegraphics[width=0.465\columnwidth]{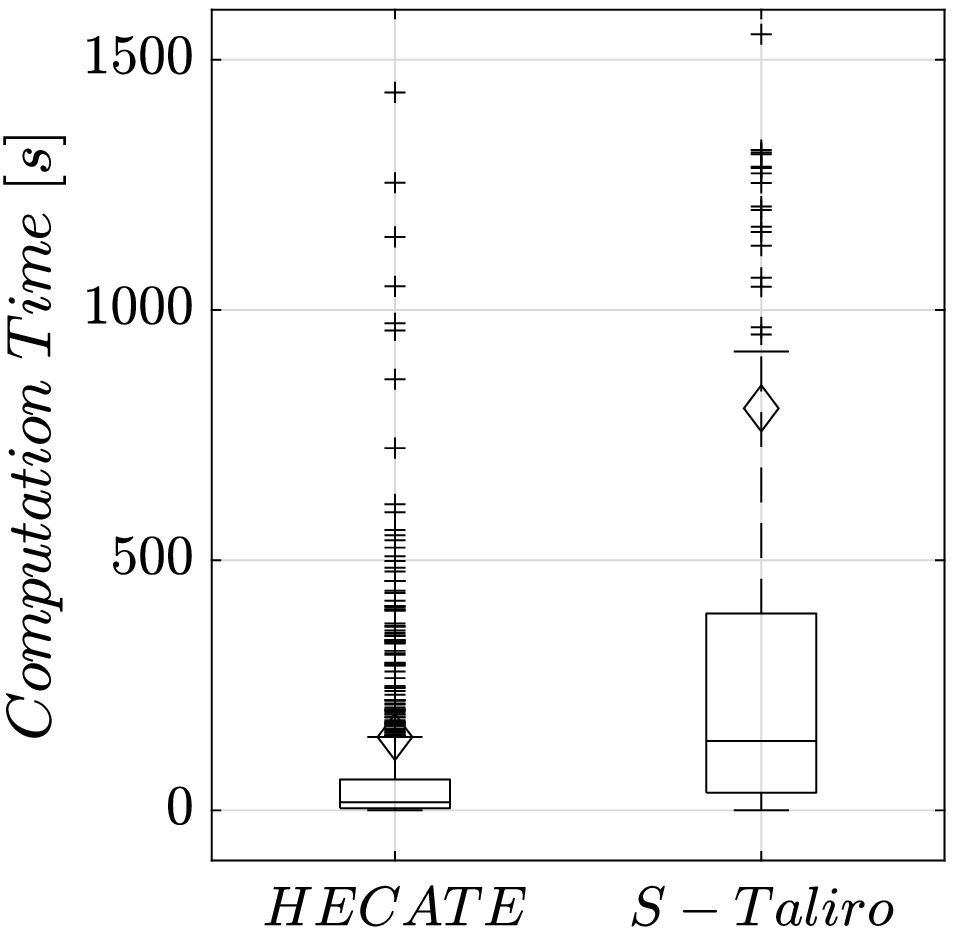}}
 \subfloat[Number of iterations\label{fig:rq2resultsiterations}]{
    \centering
    \includegraphics[width=0.465\columnwidth]{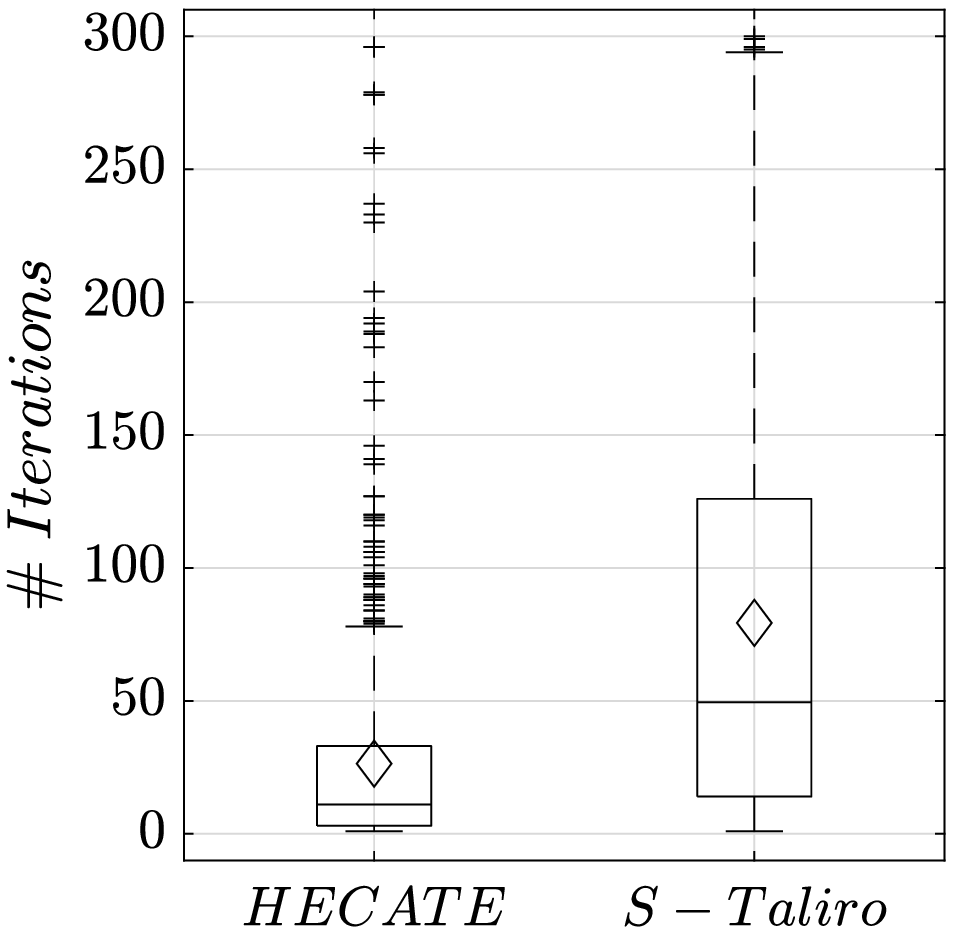}}
\caption{Time and number of iterations required by \NAME and \staliro to detect failure-revealing test cases.
    Diamonds depict the average.
    }
    \label{fig:rq2results}
\end{figure}

\textbf{Results.} 
The box plots in \Cref{fig:rq2results} report
the time and number of iterations for \NAME and \staliro.
\NAME and \staliro required on average \RQtwohecateavgtime
(\textit{min}=\RQtwohecatemintime,
\textit{max}=\RQtwohecatemaxtime,
\textit{StdDev}=\RQtwohecatestdtime).
 and \RQtwostaliroavgtime 
(\textit{min}=\RQtwostaliromintime,
\textit{max}=\RQtwostaliromaxtime,
\textit{StdDev}=\RQtwostalirostdtime) to generate the failure-revealing test cases respectively. 
They performed on average \RQtwohecateavg iterations 
(\textit{min}=\RQtwohecatemin,
\textit{max}=\RQtwohecatemax,
\textit{StdDev}=\RQtwohecatestd)
 and \RQtwostaliroavg iterations 
(\textit{min}=\RQtwostaliromin,
\textit{max}=\RQtwostaliromax,
\textit{StdDev}=\RQtwostalirostd). 
Therefore, \NAME is on average more efficient than \staliro.
It required on average
\RQtwohecatevsstalirotime ($\RQtwostaliroavgtime-
\RQtwohecateavgtime$) seconds less than \staliro.
\NAME also required on average
\RQtwohecatevsstaliro ($\RQtwostaliroavg-
\RQtwohecateavg$) less iterations than \staliro.
\Cref{tab:models} (column RQ2) reports the average time in seconds and in parenthesis the average number of iterations for each model. 
Cells with green background refer to cases in which both the time and the number of iterations of \NAME is lower than the one of \staliro. Otherwise, cells have a red background.
By considering each model separately,
the average time and number of iterations required of \NAME are both  lower than the one of \staliro for \RQtwohecatemoreefficient of the models ($13$ out of $16$). 
Wilcoxon rank sum test ($\alpha$=$0.05$, p=$2.22\cdot 10^{-3}$) confirms that \NAME is more efficient than \staliro.

\begin{Answer}[RQ2 - Efficiency]
\NAME on average required 
\RQtwohecatevsstalirotime less seconds and
\RQtwohecatevsstaliro less iterations than \staliro to generate failure-revealing test cases. It was more efficient than \staliro for \RQtwohecatemoreefficient of our benchmark models. 
\end{Answer}

\subsection{Usefulness (RQ3)}
\label{sec:rq3}
To answer RQ3, we considered a representative case study from the automotive domain. 
Our case study is a Simulink model of a hybrid-electric vehicle (HEV) developed by MathWorks for the EcoCAR Mobility Challenge~\cite{EcoCAR}. 
The HEV model has $4641$ blocks, one input port, and one output port, and has simulation time set to $400$s.
The HEV model converts electrical energy into mechanical energy. 
The model consists of different subsystems, including regenerative brake blending, battery management, power management, and software controller.
These subsystems also use 
\Simscape, \SimscapeElectrical, and \SimscapeDriveline blocks~\cite{Simscape, SimscapeElectrical, SimscapeDriveline}.
The controller is modeled as a Stateflow chart and adjusts the vehicle speed depending on the speed demand.
The HEV comes with five driving scenarios, out of which we picked the scenario Urban Cycle~3 for our experiments.

\textbf{Methodology.} 
The speed demand of Urban Cycle~3 changes within the range $[0,90]$kph. 
We created a Test Sequence representing perturbations to be applied to the speed demand.
It has four test steps, each describing the perturbation to be applied to the demanded speed within a $100$s time interval.
We added four test parameters, one for each test step.  
We set their domain to $[0,4]$kph to ensure that our perturbations are lower than 5\% of the maximum speed demand of the original driving scenario.

We designed a Test Assessment block that checks whether the delta speed, i.e., the absolute difference between the desired speed and the vehicle speed, is lower than a threshold value.
For the original speed demand of Urban Cycle~3, the delta speed was always lower than $0.5$kph.
We defined a Test Assessment block that tolerates a larger difference to account for the perturbations applied to the speed demand by the Test Sequence block.
The Test Assessment block contains three test steps.
The Test Sequence block starts applying perturbation at time instants $0$, $100$, $200$, and $300$.
The first Test Assessment step is active in the time intervals $[0,1]$, $[100,101]$, $[200,201]$, $[300,301]$. 
It does not contain any \lit{verify} or \lit{assert} statements to enable the controller to adjust the vehicle speed to the changes in the speed demand. 
The second Test Assessment step is active within the time intervals $[1,10]$, $[101,110]$, $[201,210]$, $[301,310]$ and verifies whether the delta speed is lower than $1.5$kph.
The third Test Assessment step is active within the time intervals $[10,100]$, $[110,200]$, $[210,300]$, $[310,400]$, and checks whether the delta speed is lower than the threshold value $1$kph. 
To summarize, after each perturbation to the speed demand, our Test Assessment  provides $1$s and $10$s to enable the delta speed to become lower than the threshold values $1.5$kph and $1$kph.

For this evaluation, \NAME is set to run for a maximum $300$ iterations.

\textbf{Results.}  \NAME returned a failure-revealing test case in $13$ iterations requiring $257$s ($\approx$4min). 
The values assigned to the four parameters of the Parameterized Test Sequence by the returned Test Sequence are $3.81$, $1.49$, $1.90$, and $3.66$. 
For this Test Sequence, the vehicle speed of the car ranges from $3.81$kph to $93.66$kph, while the delta speed ranges from $0$kph to $4$kph. 
The failure-revealing test case reveals the following problem. 
The controller needs more than one second to reduce the value of the delta speed below the $1.5$kph threshold for the perturbation applied to the vehicle speed by the first test step of the Test Sequence block.
We analyzed the controller behavior to detect the problem.
The controller has two modes: \text{Start\_mode}, which is entered at the beginning of the simulation, and  \text{Normal\_mode}, which regulates the engine speed.
The controller moves from \text{Start\_mode} to \text{Normal\_mode} when the engine speed exceeds 
a threshold of $800$rpm.
The original speed demand for Urban Cycle~3 was $0$kph for the first $10$s. 
Our Test Sequence applied a small perturbation to this value for its first test step ($3.81$kph). 
Due to this perturbation, the engine speed takes more than $1$s to reach $800$rpm and to bring the controller to its \text{Normal\_mode}. 
This leads to the violation of the condition specified by the test assessment, i.e., the controller needs more than one second to reduce the value of the delta speed below the $1.5$kph.

\begin{Answer}[RQ3 - Usefulness]
\NAME computed a failure-revealing test case for our automotive case study within $5$min by performing $13$ iterations. 
The failure-revealing test case shows that with a small perturbation in demanded speed, it takes the engine speed more than $1$s to reach $800$rpm and to bring the controller to its \text{Normal\_mode}. 
\end{Answer}

 \section{Discussion and Threats to Validity}
\label{sec:discussion}
This section reflects on our solution and results (\Cref{sec:reflections}) and elicits threats to validity (\Cref{sec:threats}).

\subsection{Reflections}
\label{sec:reflections}
Automatically generating failure-revealing test cases is a widely recognized software engineering problem~\cite{papadakis2019mutation,5210118}. 
Providing support for Test Sequence and Test Assessment blocks is a significant problem since these blocks are widely used by practitioners~\cite{dalton2018using,wagner2017benefits} and  support standard compliance~\cite{hoadley2010using,schmidt2014efficient}.
There is no approach that solves this problem: \NAME is the first solution addressing the problem. 
Therefore, \NAME \emph{will significantly impact}  software engineering practices by enabling the automatic generation of test cases from manually specified Test Blocks.

Our solution is \emph{novel}: search-based testing was never used with Test Sequence and Test Assessment blocks. 
The novelty concerns the development of a new testing solution addressing a relevant software engineering problem.
We reused and adapted existing solutions to solve our problem, a recommended software engineering practice.
We implemented our solution as a plugin for \staliro. 
The reuse of a mature tool (\staliro) for the development of \NAME  makes our solution easily testable, maintainable, and extensible.

We answered RQ1 and RQ2 by comparing \NAME with \staliro.
Assessing the effectiveness and efficiency of a solution by comparing it with a baseline is a \emph{sound} technique  widely used in the research literature (e.g.,~\cite{Aristeo,menghi2019generating}) also encouraged by international competitions (e.g., ARCH).
Statistical tests confirmed the significance and soundness of our results.
We answered RQ3 by assessing its applicability on a large case study.
This is a sound technique largely used in the literature (e.g.,~\cite{Aristeo}).
Finally, RQ3 assesses whether \NAME can detect failure-revealing test cases for a large case study. 
Therefore, a comparison between \NAME and \staliro is not performed since it is not the objective of RQ3.

Our results are \emph{verifiable and transparent}: our models, data, and tool are available for independent verification.

\subsection{Threats to Validity}
\label{sec:threats}

\emph{External Validity.} 
The selection of the benchmark models for RQ1 and RQ2 and the case study model for RQ3 could threaten the external validity of our results.
To mitigate this threat, we selected our benchmark models by including models
extensively used in the literature  (e.g.,~\cite{DBLP:conf/arch/ErnstABCDFFG0KM21,fehnker2004benchmarks,jin2014powertrain,Aristeo,fainekos2019robustness,Donze2010}), 
from different domains (automotive, 
ML, 
electrical, 
and aerospace), 
from industry (AFC~\cite{jin2014powertrain} is from Toyota, TUI, NNP, EU~\cite{mavridou2020ten,benchmarkLM} are from  Lockheed Martin), and with a variable number of blocks (\textit{avg}=$\approx177.3$, \textit{min}=$26$, \textit{max}=$714$, \textit{StdDev}=$\approx182.9$).
The number, size, and type of models considered in this work is consistent with similar works from this domain (e.g.,~\cite{Aristeo}).
We remark that (a)~software companies do not usually share their Simulink models, and (b)~we made our benchmark publicly available, enabling experiment replication. 
For RQ3, we selected a representative case study: it was developed by MathWorks 
for the EcoCAR Mobility Challenge~\cite{EcoCAR}.

The selection and definition of the Test Blocks and STL artifacts could threaten the external validity of our results since it influences the effectiveness and efficiency of \NAME and \staliro.
To mitigate this threat we proceeded as follows. 
For RQ1 and RQ2, we considered Test Blocks and STL artifacts defined by
(a)~including models that already included STL artifacts and manually defining Test Blocks,
(b)~including models that already included Test Blocks and manually defining STL artifacts,
(c)~including models that did not include neither Test Blocks nor STL artifacts, and
(d)~testing the equivalence of the oracle generated from the STL specification and the Test Assessment blocks.
For RQ3, we considered the model documentation for defining the Test Blocks.

Our results depend on Parameterized Test Sequence, its parameter domain, and Test Assessment blocks. 
Defining criteria to automatically define the Parameterized Test Sequence, its parameter domain, and Test Assessment blocks is out of the scope of this work and is left for future work.

Finally, the results for RQ1, RQ2, and RQ3 are likely to improve if the designers of Test Blocks are knowledgeable about the domain and have engineered the Simulink models.

\emph{Internal Validity.} 
The selection of the values of the configuration parameters of \NAME and \staliro could threaten the internal validity of our results. 
To mitigate this threat, we selected the same values for the common configuration parameters of the two tools. Therefore, our configuration does not favor any of the tools.

 \section{Related Work}
\label{sec:related}
This section presents related testing frameworks that (a)~support Simulink models, (b)~support Test Blocks, and (c)~start from manually specified test cases.

\emph{Simulink Models}. There are two main categories of testing techniques for Simulink models: the ones that rely on simulation-based software testing (e.g.,~\cite{Breach,S-Taliro,Aristeo,Falstar,Donze2010,menghi2019generating,7886937,matinnejad2016automated,10.1145/2642937.2642978,peranandam2012integrated,lindlar2010integrating,arrieta2017towards,schmidt2016model,schmidt2014efficient,pill2016simultate}), and the ones that
need to access the internal structure of the models to generate test cases. 
The techniques from the first category do not support Test Blocks (see \Cref{sec:intro}). 
\NAME extends these techniques by supporting Test Blocks and uses one of these techniques as baseline for our comparison (see \Cref{sec:evaluation}).
The techniques from the second category (e.g., Simulink Design Verifier~\cite{SimulinkDesignVerifier} and Reactis\cite{ReactisUsermanual}) need to  access the internal structure of the models to generate test cases and usually attempt to satisfy some coverage metric.  
In addition, they do not support variable-step solvers or continuous time blocks, such as the Integrator, Derivative, or Transfer Function blocks. 
Therefore, we could not use these techniques as baseline for our experiment since they do not apply to a large set of models of our benchmark.

\emph{Simulink Test Blocks}. 
A few approaches from the literature propose testing formalisms that are similar to Simulink Test Blocks~\cite{bringmann2006systematic,bringmann2007testing,srivastava2010automatic,srivastava2010automated}. However, these works do not automatically generate test cases.
Only a few approaches from the literature target Simulink Test Blocks (e.g.,~\cite{ferrer2015search,peranandam2012integrated,ayerdi2022performance}).
Unlike \NAME, 
these approaches 
(a)~do not aim to generate a failure-revealing test case, but rather aim to satisfy coverage criteria~\cite{ferrer2015search,peranandam2012integrated}, or to
identify performance-driven metamorphic relations~\cite{ayerdi2022performance}.
Lastly, some approaches target formalisms named `test sequences'~\cite{bombarda2020automata,halle2017sealtest,sheng2019constructing,halle2021test,sheng2018extended} that are not related to Simulink and are significantly different from Simulink Test Sequences.

\emph{Manually Specified Test Cases}. Several testing techniques generate new test cases from already existing, including manually defined test cases (see surveys~\cite{papadakis2019mutation,5210118,DBLP:journals/soco/KhariK19,7102580}).
However, these approaches do not target Simulink models.  \section{Conclusion}
\label{sec:conclusion}

This work presented \NAME, a testing approach for Simulink models.
Unlike existing approaches, \NAME uses the Test Sequence and Test Assessment blocks in Simulink Test to guide the search-based exploration. 
\NAME relies on Parameterized Test Sequences and on a procedure that compiles Test Assessment blocks into fitness functions.
We implemented \NAME as a plugin for \staliro. We evaluated \NAME by comparing it with a baseline.
Our results show that \NAME is more effective and efficient than our baseline. 
We also evaluated the applicability of \NAME on a representative automotive case study. 
\NAME successfully generated a failure-revealing test case for our case study.

\section*{Data Availability}
A complete replication package containing our models, evaluation data, and tool is publicly available~\cite{HECATETool}. 
The EKF model is protected by an non-disclosure agreement and can not be shared publicly.

\section*{Acknowledgment}
We acknowledge the support of the Natural Sciences and Engineering Research Council of Canada (NSERC) [funding reference numbers RGPIN-2022-04622, DGECR-2022-0040].

This research was enabled in part by support provided by Compute Ontario (\url{www.computeontario.ca}) and Compute Canada (\url{www.computecanada.ca}).

\bibliographystyle{IEEEtran}

\vfill\break
\end{document}